\documentclass[aps,showkeys,superscriptaddress,preprint,tightenlines]{revtex4}
\usepackage{times}
\usepackage{amsmath}
\usepackage{graphicx}
\usepackage{epsfig}
\usepackage{dcolumn}
\usepackage{bm}
\usepackage{color}
\usepackage{longtable}
\usepackage{array}
\usepackage{multirow}
\usepackage{orcidlink}
\newcommand{\gev}{\rm GeV}

\newcommand{\gevcs}{{\rm GeV}/c^2}
\newcommand{\mev}{\rm MeV}
\newcommand{\mevcs}{{\rm MeV}/c^2}

\newcommand{\pbv}{{\rm pb}^{-1}}
\newcommand{\fbv}{{\rm fb}^{-1}}

\newcommand{\xyz}{\rm XYZ}

\newcommand{\zc}{Z_c(3900)}
\newcommand{\zcp}{Z_c(4020)}
\newcommand{\zcs}{Z_{cs}(3985)}

\newcommand{\hc}{h_c}
\newcommand{\pphc}{\pi^+\pi^- h_c}

\newcommand{\psp}{\psi(2S)}
\newcommand{\psip}{\psi(2S)}
\newcommand{\pspp}{\psi(3770)}
\newcommand{\jpsi}{J/\psi}

\newcommand{\EE}{e^+e^-}

\newcommand{\pp}{\pi^+\pi^-}

\newcommand{\ccb}{c\bar{c}}

\newcommand{\ppjpsi}{\pi^+\pi^- J/\psi}

\newcommand{\beq}{\begin{equation}}
\newcommand{\eeq}{\end{equation}}
\newcommand{\bitm}{\begin{itemize}}
\newcommand{\eitm}{\end{itemize}}
\begin{document}

\title{ Discovery of XYZ particles at the BESIII Experiment~\footnote{Talk at the International Symposium of ``50 Years Discovery of the J Particle'', Beijing, October 20, 2024.} }
\author{Chang-Zheng Yuan~\orcidlink{0000-0002-1652-6686}}
 \email{yuancz@ihep.ac.cn}
 \affiliation{Institute of High Energy Physics, Chinese Academy of Sciences,
 Beijing 100049, China}
 \affiliation{University of Chinese Academy of Sciences, Beijing 100049, China}

\begin{abstract}

Charmonium is a bound state of a charmed quark and a charmed
antiquark, and a charmoniumlike state is a resonant
structure that contains at least a charmed quark-antiquark pair 
but has properties that are incompatible with a conventional charmonium
state. The charmoniumlike states are also called $\xyz$ particles
to indicate their underlying nature is still unclear. 
The BESIII experiment has contributed significantly in the
study of the $\xyz$ particles, and here we review the discoveries 
of the $\zc$, $\zcp$, and $\zcs$ tetraquark states
and the observations of several new vector charmoniumlike states at
the BESIII experiment.

\end{abstract}


\maketitle

\section{Introduction}\label{Sec:intro}

In the conventional quark model, mesons are comprised of a quark
and antiquark pair, while baryons are comprised of three quarks.
A bound state of a charmed quark ($c$) and a charmed antiquark
($\bar{c}$) is named charmonium. The first charmonium state, the
$\jpsi$, was discovered at BNL~\cite{ting} and at
SLAC~\cite{richter} in 1974, and since then, all the charmonium
states below the open-charm threshold and a few vector charmonium
states above the open-charm threshold have been established~\cite{pdg}.

In addition to the charmonium states, almost all of the hadrons
that have been observed to date, including three-quark
baryons and other quark-antiquark mesons~\cite{pdg}, can be
described well by QCD~\cite{Brambilla:2004jw,review4,Brambilla:2014jmp} 
and QCD-inspired potential models~\cite{eichten,godfrey,barnes}.
Exotic hadronic states with configurations not limited to two or three
quarks have been the subject of numerous theoretical proposals and
experimental searches~\cite{Jaffe:2004ph,klempt}. These proposed exotic 
hadrons include hadron-hadron molecules, diquark-diantiquark tetraquark 
states, hadro-quarkonia, quark-antiquark-gluon hybrids, multi-gluon 
glueballs, and pentaquark baryons.

Many $\xyz$ states were discovered at the BaBar~\cite{babar} 
and Belle~\cite{belle} $B$-factories during the first decade 
of this century~\cite{PBFB}, and some of them are good candidates
of exotic hadrons. 
Usually we use $Z_c({\rm xxxx})$ to denote a charmoniumlike 
state with mass roughly ${\rm xxxx}$~MeV/$c^2$ that contains 
a heavy quark pair $c\bar{c}$ and with non-zero isospin; 
$Y({\rm xxxx})$ for a vector charmoniumlike
state (called $\psi({\rm xxxx})$ by PDG~\cite{pdg}), and $X({\rm
xxxx})$ for states with other quantum numbers.

Although the BaBar and Belle experiments
finished data taking in 2008 and 2010, respectively, their data are
still used for various physics analyses. In 2008, two new
experiments: BESIII~\cite{bes3}, a $\tau$-charm factory experiment
at the BEPCII $\EE$ collider; and LHCb~\cite{lhcb_detector}, a
$B$-factory experiment at the LHC $pp$ collider, started data
taking, and have been contributing to the study of charmonium and
charmoniumlike states ever since. Detailed discussions of the
experimental observations and the theoretical interpretation can
be found in many good reviews~\cite{review1,review2,review3,review4,
Olsen:2017bmm,Wang:2025sic,Esposito:2016noz,Lebed:2016hpi}. 

The BESIII experiment at the BEPCII double ring $\EE$ collider
observed its first collisions in the $\tau$-charm energy region in
July 2008. After a few years of
running at center-of-mass (c.m.) energies for its well-defined
physics programs~\cite{BESIII_YB}, i.e., at the $\jpsi$ and $\psp$
peaks in 2009 and the $\pspp$ peak in 2010 and 2011, the BESIII
experiment began to collect data for the study of the $\xyz$
particles~\cite{BESIII_YB}. The first data sample was collected at the
$\psi(4040)$ resonance in May 2011 with an integrated luminosity
of about $0.5~\fbv$. 

In summer 2012, the LINAC of the BEPCII was upgraded and 
made it possible for BESIII experiment to collect data at c.m. 
energies up to $4.6~\gev$. A data sample of $525~\pbv$ was collected 
at $4.26~\gev$ from December 14, 2012 to January 14, 2013,
with which the $\zc$ charged charmoniumlike state was
discovered~\cite{zc3900}. This observation had considerable impact
on the subsequent running schedule of the experiment: more data
between $4.13$ and $4.60~\gev$ dedicated to the $\xyz$
related analyses were recorded~\cite{lum_songwm}. The highest beam
energy was further increased from $2.3$ to $2.5~\gev$ in summer
2019, making it possible to collect data at c.m. energies up to 
$5.0~\gev$.

The BESIII experiment has collected $e^+e^-$ collision data across 
a c.m. energy range from $1.84$ to $4.95~\mathrm{GeV}$ until now. 
The data samples used for the $\xyz$ study cover the energy range
between $4.0$ and $5.0$~GeV, with a typical integrated luminosity of
$500~\pbv$ at each energy point. Data samples with an integrated 
luminosity of $826~\pbv$ at 104 energy points between $3.8$ 
and $4.6~\gev$~\cite{lum_Rscan} was also used for the $\xyz$ study.
These data sets include 199 energy points with a total integrated 
luminosity of $26$~fb$^{-1}$.

In this article, we review the discoveries of the $\zc$, $\zcp$, and 
$\zcs$ tetraquark states and the observations of several new vector 
charmoniumlike states. More BESIII results can be found in 
Refs.~\cite{Yuan:2021wpg,Mitchell:2020omp,BESIII:2020nme,Yuan:2018inv,Yuan:2015kya}.

\section{\boldmath Discovery of $\zc$, $\zcp$, and $\zcs$ tetraquark states}
\label{Sec:z}

Searching for charged charmoniumlike states is one of the most
promising ways of establishing the existence of the exotic
hadrons, since such a state must contain at least four quarks and,
thus, could not be a conventional meson. These searches have been
concentrated on decay final states that contain one charged pion
and a charmonium state, such as the $\jpsi$, $\psp$, and $\hc$,
since they are narrow and their experimental identification is
relatively unambiguous.

\begin{figure}[htbp]
\begin{center}
\includegraphics[height=5.0cm]{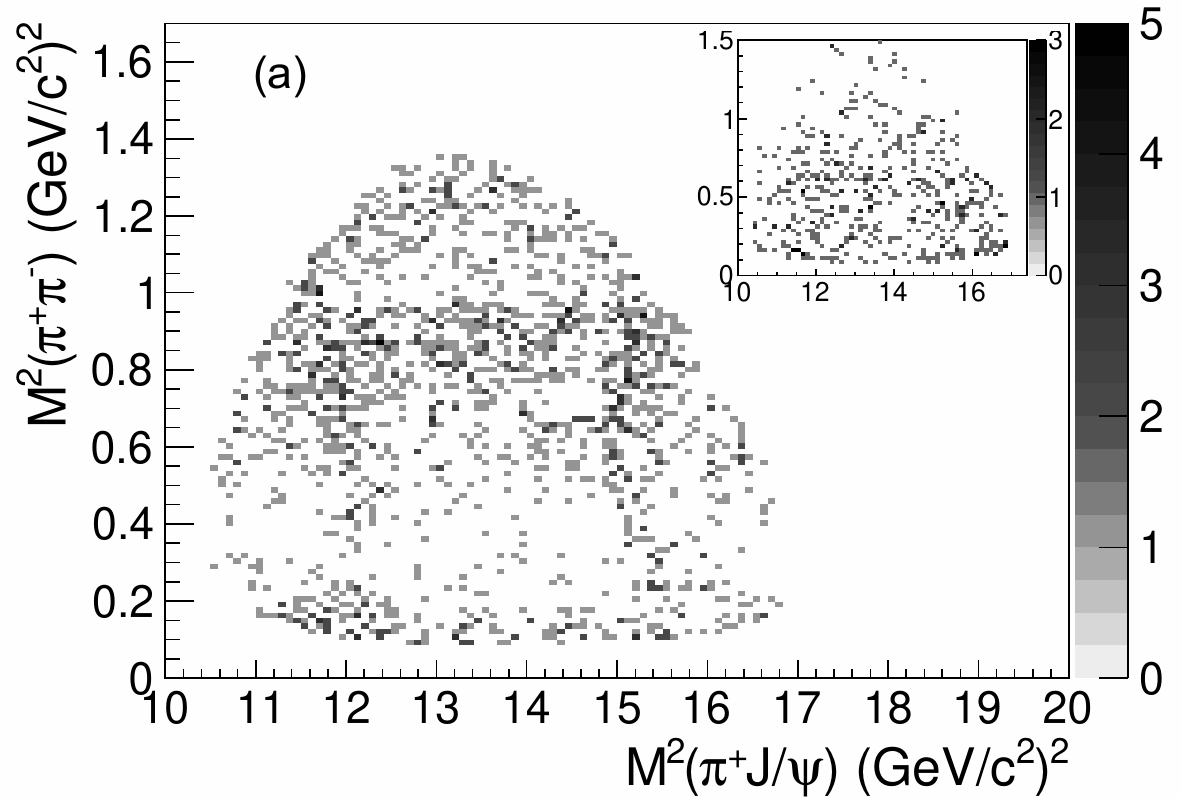}
\includegraphics[height=5.0cm]{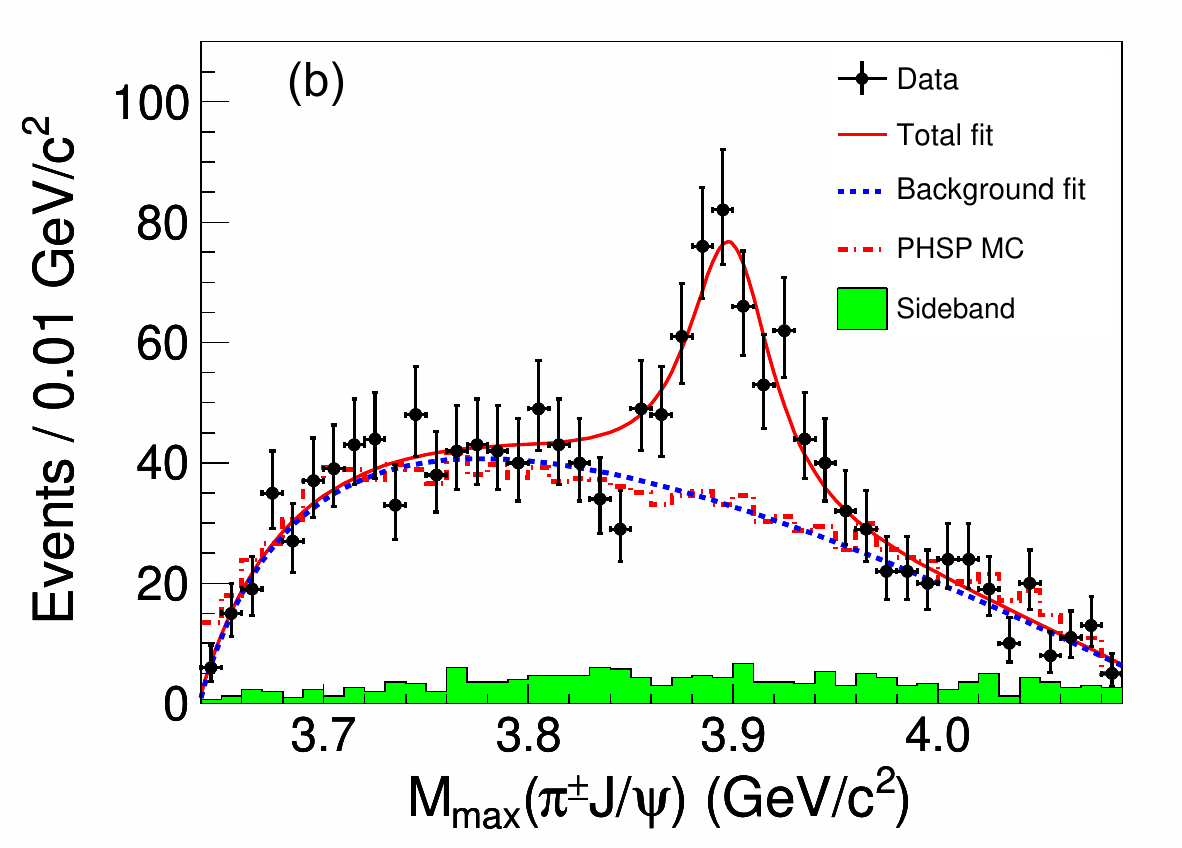}
\caption{Dalitz plot for selected $\EE\to \ppjpsi$ events in the
$\jpsi$ signal region (a, the inset show background events from
the $\jpsi$ mass sidebands) and the $\zc$ signal in the
$M_{\mathrm{max}}(\pi J/\psi)$ (b)~\cite{zc3900}. Points with
error bars are data, the curves are the best fit, the dashed
histograms are the phase space distributions and the shaded
histograms are the non-$\ppjpsi$ background estimated from the
normalized $\jpsi$ sidebands.} \label{zc}
\end{center}
\end{figure}

The BESIII experiment studied the $\EE\to \ppjpsi$ process using a
$525~\pbv$ data sample at a c.m. energy of
$4.26~\gev$ in 2013~\cite{zc3900}. About $1500$ signal events were
observed and the cross section was measured to be $(62.9\pm 1.9
\pm 3.7)$~pb. The intermediate states in this three-body system
were studied by examining the Dalitz plot of the selected
candidate events, as shown in Fig.~\ref{zc}.

In addition to the known $f_0(500)$ and $f_0(980)$ structures in
the $\pp$ system, a structure at around $3.9~\gevcs$ was observed
in the $\pi^\pm \jpsi$ invariant mass distribution with a
statistical significance larger than $8\sigma$, which is referred
to as the $\zc$. A fit to the $\pi^\pm\jpsi$ invariant mass
spectrum (see Fig.~\ref{zc}) determined its mass to be $(3899.0\pm
3.6\pm 4.9)~\mevcs$ and its width $(46\pm 10\pm 20)~\mev$.

A measurement performed at the Belle experiment that was released
subsequent to the BESIII paper reported the observation of the
$\zc$ state (referred to as $Z(3900)^+$ in the Belle paper)
produced via the initial state radiation (ISR) process with 
a mass of $(3894.5\pm 6.6\pm 4.5)~\mevcs$ and 
a width of $(63\pm 24\pm 26)~\mev$ with 
a statistical significance larger than $5.2\sigma$~\cite{belle_zc}. 
These observations were later
confirmed by an analysis of CLEO-c data at a c.m. energy of
$4.17~\gev$~\cite{seth_zc}, with a mass and width that agree with
the BESIII and Belle measurements. The $\zc$ is thus the first 
confirmed tetraquark state, and its spin-parity quantum numbers 
are measured as $J^P=1^+$~\cite{zc3900_jpc} and 
its isospin $I=1$. 

\begin{figure}[htbp]
\begin{center}
\includegraphics[width=0.45\textwidth]{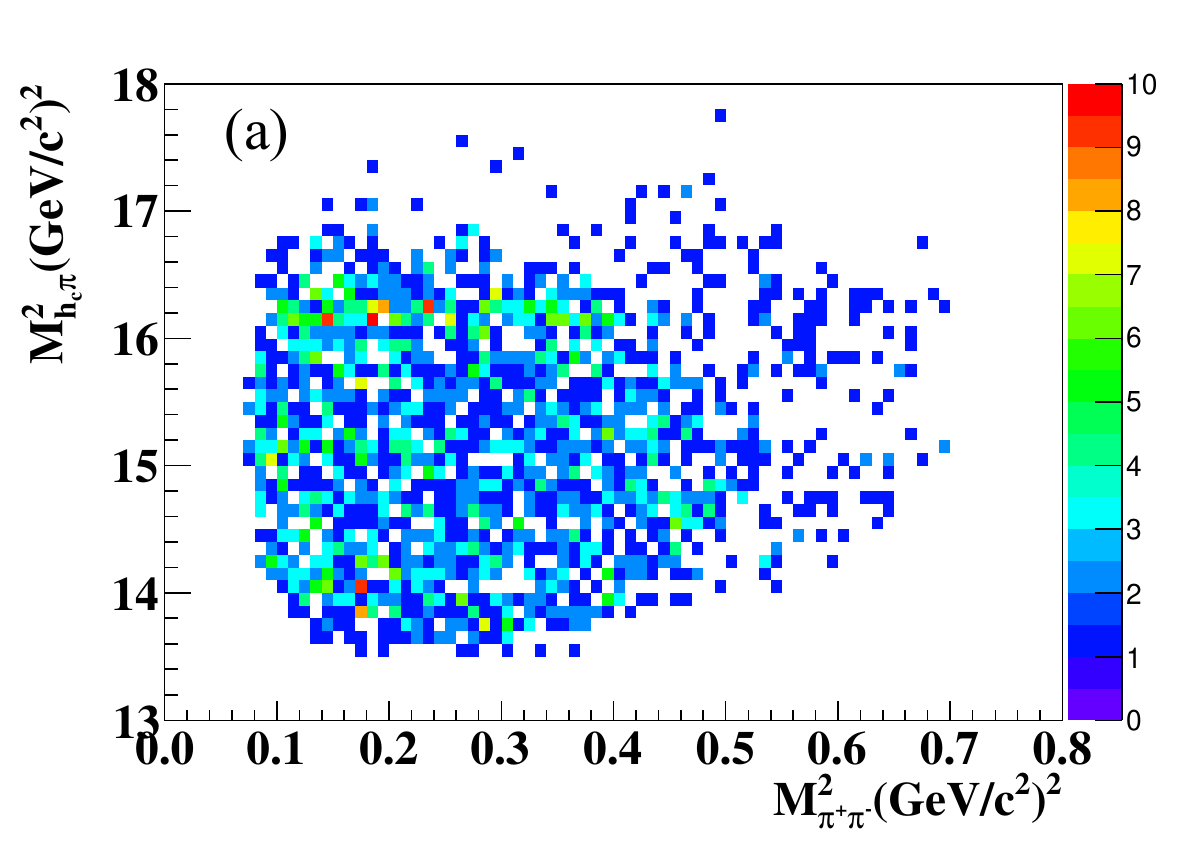}
\includegraphics[width=0.45\textwidth]{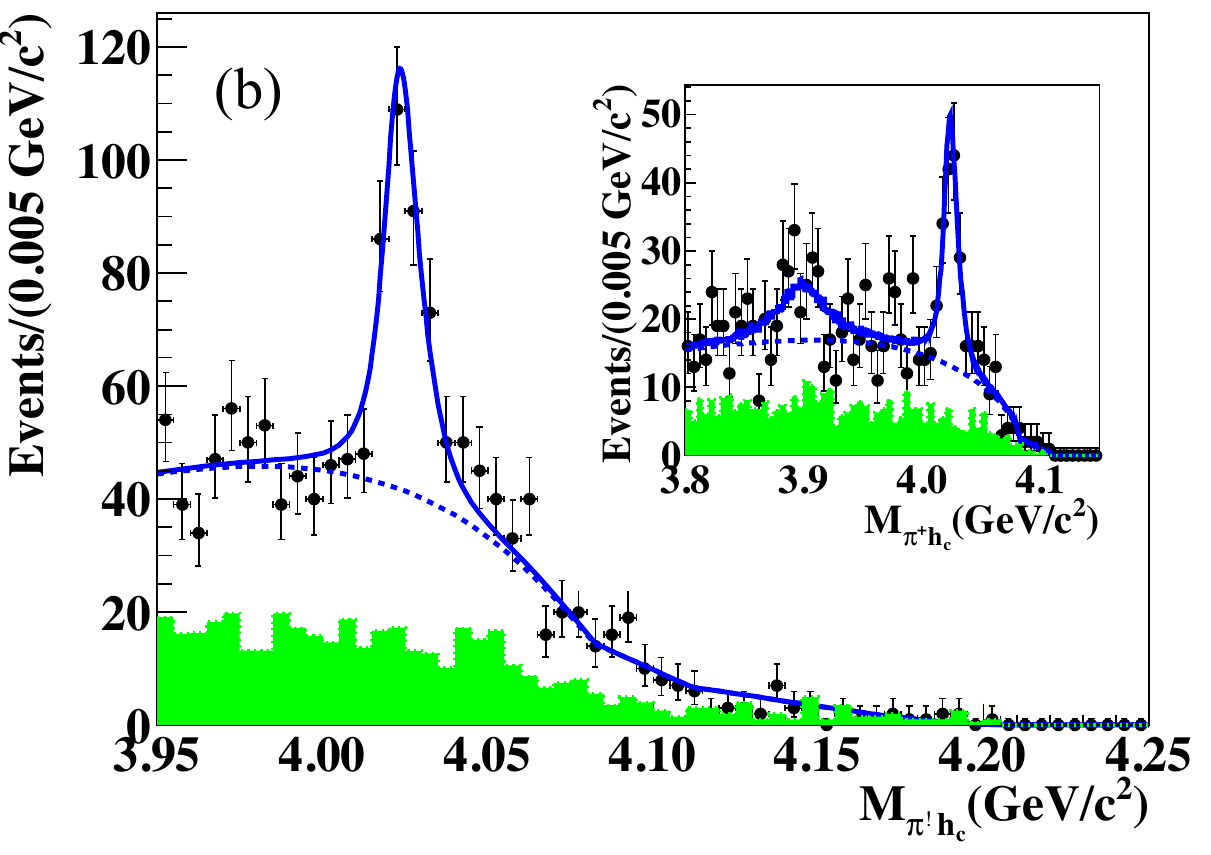}
\caption{Dalitz plot ($M^2_{\pi^+\hc}$ vs. $M^2_{\pp}$) for
selected $\EE\to \pphc$ events (a) and the $\zcp$ signal observed
in $\pi\hc$ invariant mass spectrum (b)~\cite{zc4020}. Points with
error bars are data, the solid curves are the best fit, the shaded
histograms are the non-$\pphc$ background estimated from the
normalized $\hc$ sidebands.} \label{zcp}
\end{center}
\end{figure}

The process $\EE\to \pphc$ was observed at c.m. energies of
$3.90-4.42~\gev$~\cite{zc4020}.
Intermediate states of this three-body system were studied by
examining the Dalitz plot of the selected $\pphc$ candidate
events. There are no clear structures in the $\pp$ system, 
surprisingly, there is distinct evidence for an exotic
charmoniumlike structure in the $\pi^\pm\hc$ system, as clearly
evident in the Dalitz plot shown in Fig.~\ref{zcp}. This figure
also shows projections of the $M(\pi^\pm\hc)$ (two entries per
event) distribution for the signal events as well as the
background events estimated from normalized $\hc$ mass sidebands.
There is a significant peak at around $4.02~\gevcs$ (the $\zcp$),
and there are also some events at around $3.9~\gevcs$ that could
be due to the $\zc$. The mass and width of the $\zcp$ were
measured to be $(4022.9\pm 0.8\pm 2.7)~\mevcs$ and $(7.9\pm 2.7\pm
2.6)~\mev$, respectively. The statistical significance of the
$\zcp$ signal is greater than $8.9\sigma$.

The minimal quark content of the charged charmoniumlike 
states $\zc$ and $\zcp$ are $\ccb u\bar{d}$. In 2020, BESIII 
experiment discovered a $\zcs$ state with quark content 
$\ccb u \bar{s}$ in the $K^-$ recoil-mass spectra in 
$\EE\to K^-(D^+_s\bar{D}^{*0}+D^{*+}_s\bar{D}^0)$~\cite{BESIII:2020qkh} 
with a mass of 3983~MeV/$c^2$ and a width of about 10~MeV and found 
evidence for its neutral partner~\cite{BESIII:2022qzr}.
These indicate that the $\zcs$ states form isospin doublet. 

\begin{figure*}[htbp]
\centering
  \includegraphics[width=0.8\textwidth]{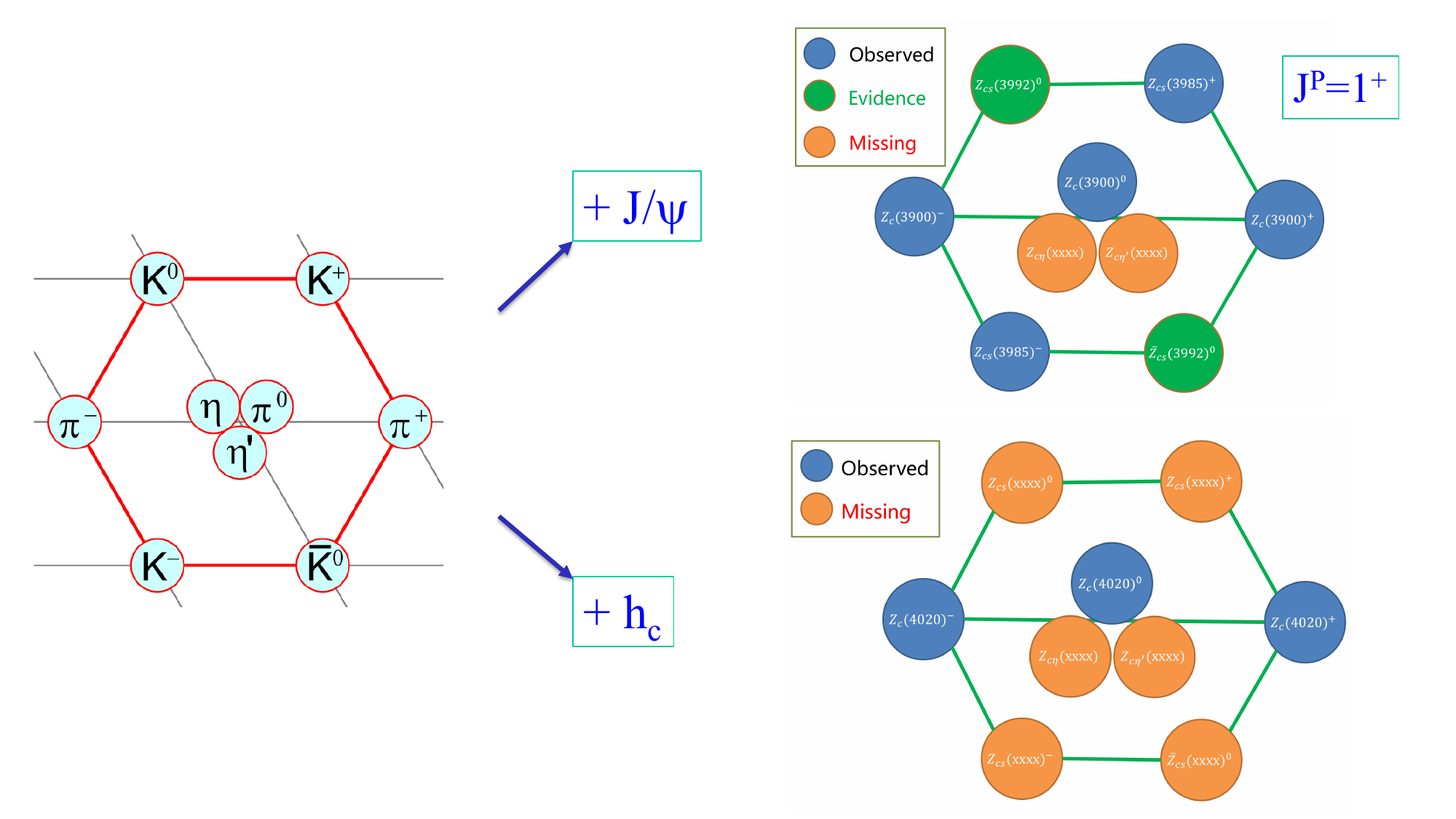}
\caption{The possible multiplets of the $Z_c(3900)$ and $Z_c(4020)$.}\label{fig:zc}
\end{figure*}

The $Z_c(3900)$ ($Z_c(4020)$) and $Z_{cs}$ states may form multiplets 
shown in Fig.~\ref{fig:zc}, the missing states can be searched for 
with the existing or future data samples.

\section{\boldmath Discovery of new vector charmoniumlike $Y$ states}\label{Sec:y}

A major advantage of $e^+e^-$ colliders is the ability to perform
energy scan, allowing the study of the vector charmonium $\psi$ and charmoniumlike
$Y$ states through cross section line shapes. This approach has been 
applied at the BESIII experiment using the fine scan data sets including 
199 energy points with a total integrated luminosity of 26~fb$^{-1}$.
BESIII measured cross sections of a large number of exclusive
processes, covering hidden-charm, open-charm, and light hadron final states.

The $Y$ states were first discovered in the ISR processes in the $B$-factory
experiments~\cite{babar_y4260,Belle:2007dxy,belle_y4660,BaBar:2006ait}, 
and they have $J^{PC}=1^{--}$. So these state can also be produced 
directly in $\EE$ annihilation experiment. Much improved measurements of 
the $Y(4260)$, $Y(4360)$, and $Y(4660)$ are achieved, and new 
vector charmoniumlike $Y$ states are observed at BESIII.

\begin{figure}[htb]
    \centering
    \includegraphics[width=0.95\textwidth]{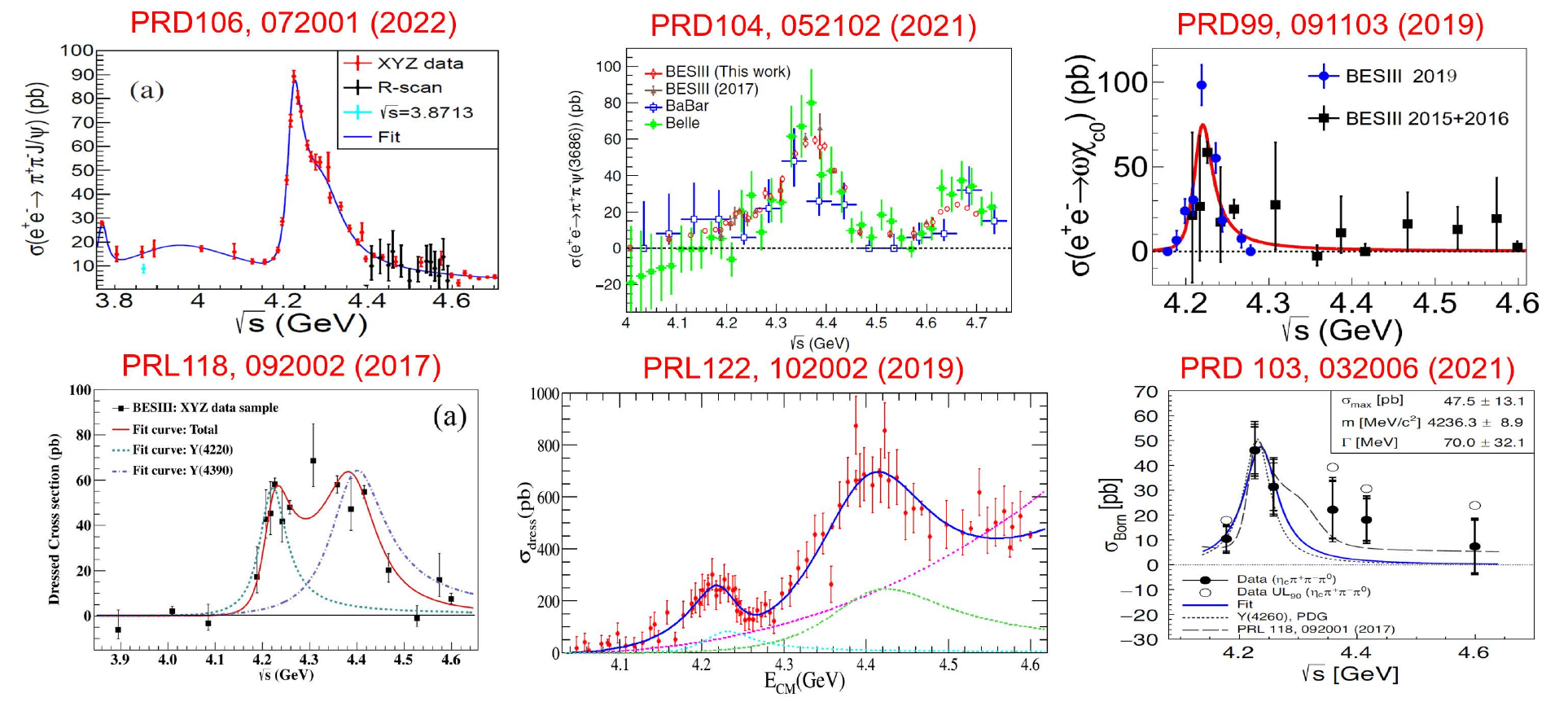}
    \caption{The $Y(4230)$ signals observed in the BESIII data.}
    \label{Y4230_BESIII}
\end{figure}

The most precise measurements of the $Y(4260)$ are from the BESIII
experiment~\cite{BESIII:2016bnd,BESIII:2022qal}. By doing a high
luminosity energy scan in the vicinity of the $Y(4260)$,
BESIII found the peak of the $Y(4260)$ is much lower (so it is now 
named the $\psi(4230)$) than that from
previous measurements and the width is narrower, and there is a high
mass shoulder with a mass of $4.32$~GeV/$c^2$ if fitted with a BW
function. Since then, more new decay modes of the $\psi(4230)$ were
observed (see Fig.~\ref{Y4230_BESIII}) including
$\pphc$\cite{BESIII:2016adj,BESIII:2025bce}, 
$\pp\psp$~\cite{BESIII:2021njb}, 
$\omega\chi_{c0}$~\cite{BESIII:2019gjc}, 
$\pi \bar{D} D^*+c.c.$~\cite{BESIII:2018iea}, 
$\pi \bar{D}^* D^*$~\cite{BESIII:2023cmv}, and 
$K\bar{K}\jpsi$~\cite{BESIII:2022joj,BESIII:2023wqy}.

The cross sections of $e^+e^-\rightarrow K^+K^-J/\psi$ at
c.m. energies from $4.127$ to $4.950$~GeV are 
measured~\cite{BESIII:2022joj,BESIII:2023wqy}. 
Three resonant structures are
observed in the line shape of the cross sections. The mass and
width of the first structure are measured to be
$(4225.3\pm2.3\pm21.5)$~MeV/$c^2$ and ($72.9\pm6.1\pm30.8$)~MeV,
respectively. They are consistent with those of the established
$\psi(4230)$. The second structure is observed for the first time
with a statistical significance greater than $8\sigma$, denoted as
$Y(4500)$. Its mass and width are determined to be
$(4484.7\pm13.3\pm24.1)$~MeV/$c^2$ and $(111.1\pm30.1\pm15.2)$~MeV,
respectively. The third structure is observed for the first time 
with a mass of $(4708_{-15}^{+17}\pm21)$~MeV/$c^2$ and a width of 
$(126_{-23}^{+27}\pm30)$~MeV with a significance over $5\sigma$, 
denoted as $Y(4710)$. 

With the world's largest $e^+e^-$ scan data sample
between $4.226$ and $4.950$~GeV accumulated by BESIII, the Born
cross sections of $e^+e^-\to D_s^{\ast+}D_s^{\ast-}$ are 
measured precisely~\cite{BESIII:2023wsc}. 
Besides two enhancements in the energy dependent cross sections at 
around $4.2$ and $4.45$~GeV/$c^2$ that may come from the $\psi(4160)$ or $\psi(4230)$ 
and the $\psi(4415)$, respectively, a third resonance structure ($Y(4790)$) 
is observed at around $4.7$$\sim$$4.8$~GeV/$c^2$ with statistical significance 
greater than $6.1\sigma$. Due to the limited number of data
points around $4.79$~GeV, the fitted mass of the third structure
varies from $4786$ to $4793$~MeV/$c^2$ and the width from $27$ to $60$~MeV. 

\begin{figure}[htb]
    \centering
    \includegraphics[width=0.95\textwidth]{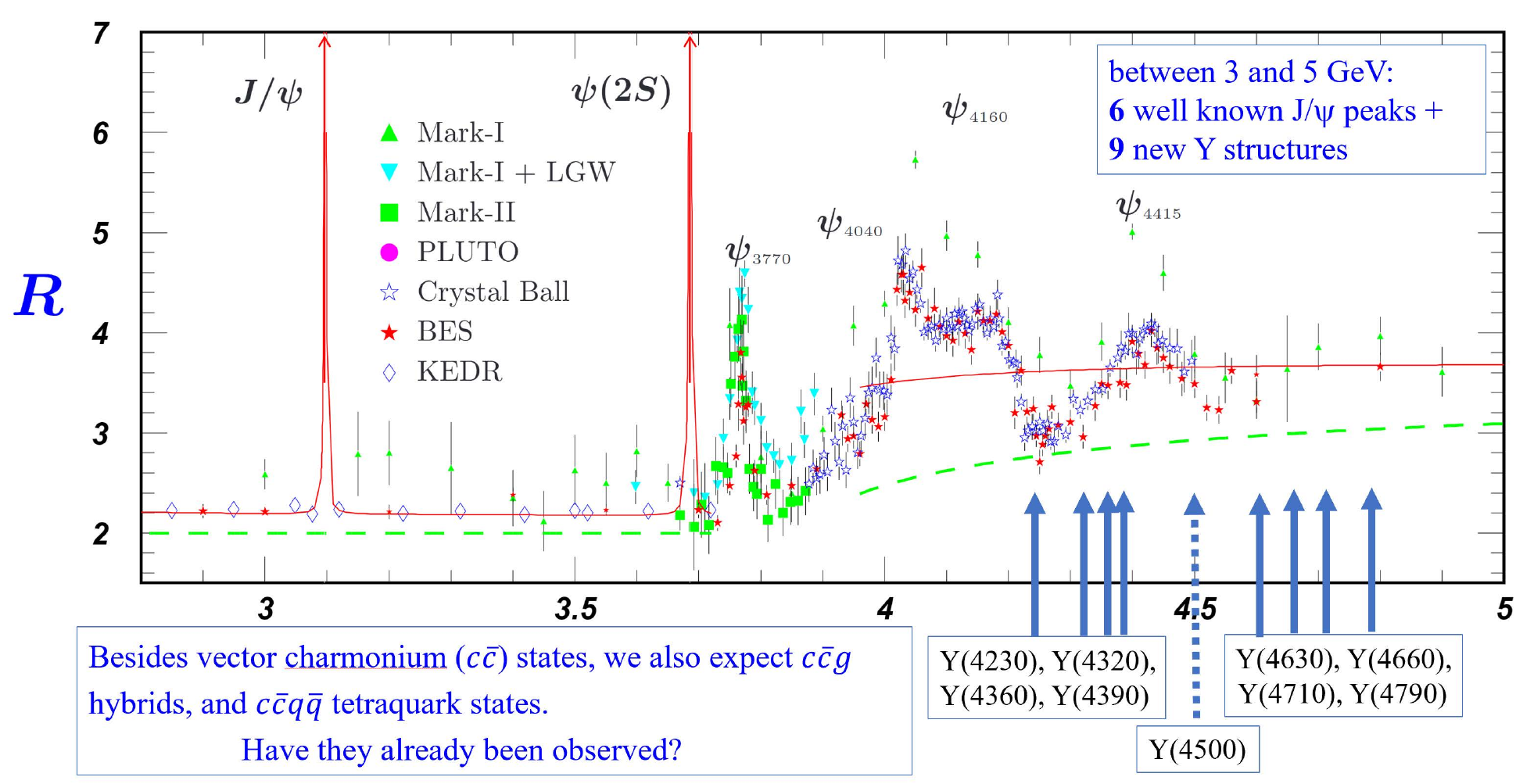}
\caption{Supernumerary vectors in the charmonium mass region may suggest 
existence of exotic states.}
    \label{Super_BESIII}
\end{figure}

In the charmonium energy region between $3$ and $5$~GeV, 
we now have identified 6 well known $\psi$ peaks 
($\jpsi$, $\psip$, $\pspp$, $\psi(4040)$, $\psi(4160)$, 
and $\psi(4415)$) and 9 new $Y$ structures ($Y(4230)$, $Y(4320)$, 
$Y(4360)$, $Y(4390)$, $Y(4500)$, $Y(4630)$, $Y(4660)$, 
$Y(4710)$, and $Y(4790)$), as indicated in Fig.~\ref{Super_BESIII}. 
They are all vector states and they cannot be all charmonium 
states~\cite{Deng:2023mza}. While more experimental 
efforts are needed to resolve the origins of these states, 
theoretical efforts are also necessary to identify if the 
vector charmonium hybrids and/or tetraquark states
have already been observed.

\begin{figure}[htb]
    \centering
    \includegraphics[width=0.95\textwidth]{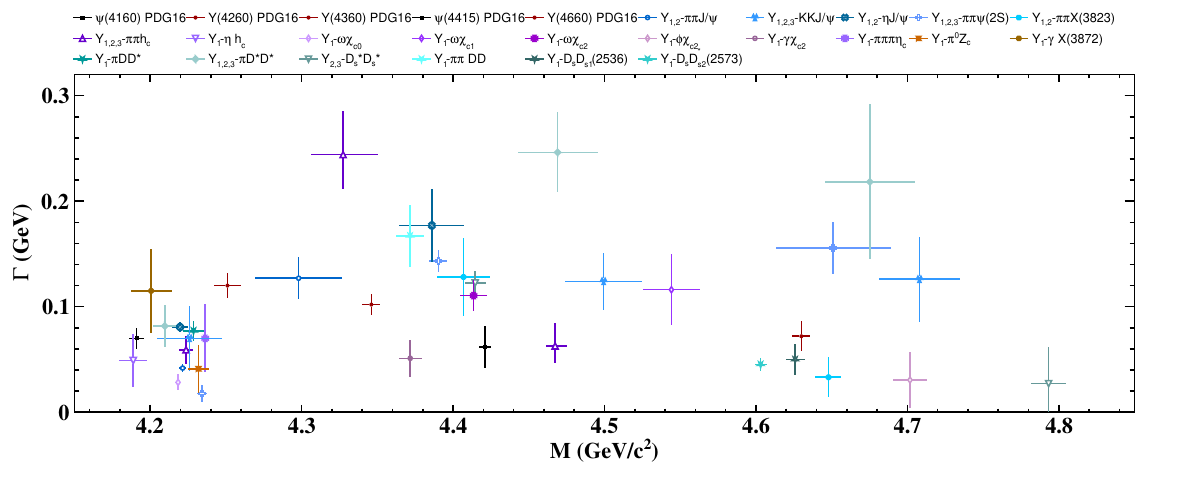}
    \caption{The parameters of the vector states obtained from single-channel analyses~\cite{Guo:2025ady}.}
    \label{Ys}
\end{figure}

The resonance parameters of the $Y$ states are typically
determined by fitting the cross section of the exclusive process
using a model that includes Breit-Wigner functions for the
resonant structures and a power-law term $1/s^{n}$ for the
continuum contribution~\cite{Guo:2025ady}. 
The number of resonant structures included in the fit
is guided by a hypothesis test, with additional resonances
incorporated only if their statistical significance exceeds
$5\sigma$. The extracted mass and width of the $Y$ states 
from each exclusive process are summarized in Fig.~\ref{Ys}.

In the vicinity of the $Y(4230)$, resonance parameters extracted
from different processes are relatively consistent. At
higher c.m. energies, however, the parameters of these $Y$ states vary
significantly from channel to channel. This inconsistency is likely due to
the limitations of the current cross-section-fit strategy, which
generally neglects coupled-channel effects that may significantly 
influence the results~\cite{Eichten:1979ms}. A more rigorous
coupled-channel analysis such as those proposed in
Refs.~\cite{Husken:2024hmi,Salnikov:2024wah,Lin:2024qcq,Nakamura:2023obk}, 
incorporating precise measurements of
two-body (and possibly three-body and four-body) open-charm cross
sections as well, would offer a more robust and comprehensive 
understanding of the $Y$ spectrum---though it presents
considerable theoretical and experimental challenges.

\section{Summary and Prospects}
\label{Sec:Summary}

With the capability of adjusting the $\EE$ c.m. energy to the
peaks of resonances, combined with the clean experimental
environments due to near-threshold operation, BESIII is uniquely
able to perform a broad range of critical measurements of
charmonium physics, and the production and decays of many of the
nonstandard $\xyz$ particles. 

BESIII discovered the $\zc$ tetraquark state which is the first
confirmed charged charmoniumlike state and 
``opened fresh vista on matter''~\cite{vista}, followed by
discovering its siblings $\zcp$ and $\zcs$. These findings point to  
the existence of a rich spectrum of tetraquark states. Furthermore, BESIII
observed several new vector states in final states with a charmed 
quark-antiquark pair. Not all of these states can be 
accommodated within the conventional quark model description of charmonium,
suggesting that some of them may be exotic states such as hadronic molecules,
tetraquark states, or charmonium hybrids.

Since delivering its first physics data in 2009, BESIII has accumulated 
more than $35$~fb$^{-1}$ of integrated luminosity 
across c.m. energies from $1.84$ to $4.95$~GeV. 
A major upgrade in 2024 significantly extended the collider's 
capabilities: the maximum beam energy of BEPCII was raised to $2.8$~GeV, 
pushing the c.m. energy reach to $5.6$~GeV and opening a new energy frontier.
Concurrently, the peak luminosity was increased by a factor of three 
for c.m. energies between $4.0$ and $5.6$~GeV.

The data from the upgraded collider, to be collected over the next decade, 
together with those collected before, will enable a comprehensive 
research program. This includes detailed 
studies of the $\xyz$ states, completing tetraquark multiplets, 
searching for higher-mass vector charmonium and charmoniumlike 
$Y$ states, and ultimately discovering new physical rules and 
new phenomena~\cite{BESIII:2022mxl}. 

\section*{Acknowledgments} 
This work is supported in part by National Key Research 
and Development Program of China under Contract No.~2020YFA0406300, 
and National Natural Science Foundation of China (NSFC) under contract 
No.~12361141819.

\end{document}